\documentclass[prd, twocolumn,preprintnumbers, showpacs, groupedaddress]{revtex4}
\usepackage{graphicx}
\usepackage[mathlines]{lineno}
\usepackage{amsmath}
\renewcommand{\vec}[1]{\mathbf{#1}}
\usepackage{slashed}
\usepackage{natbib}
\usepackage{bm}
\usepackage{amsfonts}
\usepackage{amssymb}
\usepackage{booktabs}
\usepackage{epsfig}
\usepackage{subfigure}
\usepackage{hyperref}
\hypersetup{
      colorlinks=true,
      citecolor=blue,
      linkcolor=blue,
      urlcolor=blue}
\everymath{\displaystyle}
\usepackage[usenames,dvipsnames]{color}
\usepackage{bm}
\newcommand{\bea}{\begin{eqnarray}}
\newcommand{\eea}{\end{eqnarray}}

\begin{document}
\title{Primordial magnetic field and kinetic theory with Berry curvature}
\author{Jitesh R. Bhatt$^{1}$}
 \email[e-mail:]{jeet@prl.res.in}
\author{Arun Kumar Pandey$^{1, 2}$}
\email[e-mail:]{arunp@prl.res.in, \\ arun_pandey@iitgn.ac.in}
\affiliation{%
\centerline{$^{1}$ Physical Research Laboratory, Theory Division, Ahmedabad 380 009, India}\\
\centerline{$^{2}$ Department of Physics, Indian Institute of Technology, Gandhinagar, Ahmedabad 382 424, India}
}
 \bigskip
\begin{abstract}
We study generation of magnetic field in primordial plasma of the standard model (SM) particles at
temperature $T>80$~TeV much higher than the electroweak scale. It is assumed that there is an excess number of
right-handed electrons over left-handed positrons in the plasma. Using the Berry-curvature modified kinetic 
theory to incorporate effect of the Abelian anomaly, we show that this chiral-imbalance leads to generation
of hyper-magnetic field in the plasma in both the collision dominated and the collisionless regimes. It is 
shown that in the collision dominated regime the chiral-vorticity effect can generate finite vorticity in the 
plasma together with the magnetic field. Typical strength of the generated magnetic field is $10^{27}$~Gauss
at $T\sim 80$~TeV with the length scale $10^5/T$ whereas the Hubble length scale is $10^{13}/T$. Further the
instability can also generate the magnetic field of order $10^{31}$~Gauss at typical length scale $10/T$.
But there may not be any vorticity generation in this regime. We show that the estimated values of magnetic
filed are consistent with the bounds obtained from present observations. 
\end{abstract}
\maketitle
\section*{Introduction}
There is a strong possibility that the observed magnetic fields in galaxies and in inter-galactic medium could 
be due to some process in the very early Universe. Understanding the origin and dynamics of the primordial 
magnetic field is one of the most intriguing problems of the cosmology (see the recent reviews\cite{Kandus:2010a,
Widrow:2011hs,Yamazaki:2012ab}). It should be noted here that there still exists a possibility that the fields
may not be of primordial origin but might be created during the gravitational collapse of galaxies
\cite{Kulsrud:1996b,Schleicher:2011abc}.
In the present work we are interested in the primordial origin of the magnetic fields. There exist several models
describing the generation of primordial magnetic fields in terms of cosmological defects\cite{Vilenkin_91,
Vachaspati_91a,Vilenkin_97}; phase transitions\cite{Vachaspati_91b,Enqvist_93,Kibble_95,Quashnock_89};
inflation\cite{Carroll_90,Giovannini_00a}; electroweak Abelian anomaly\cite{Joyce_97,Cornwall_97};
string cosmology\cite{Gasperini_95,Lemoine_95}; temporary electric charge nonconservation\cite{Dolgov_93a};
trace anomaly\cite{Dolgov_93b} or breaking gauge invariance\cite{Turner_88}. In a recent work\cite{Naoz_13}, 
it was shown that the process like Biermann battery can play a role in generating the primordial magnetic field
just after the recombination era.

In recent times there has been a considerable interest in studying role of the quantum chiral anomaly in 
generation of the primordial magnetic field \cite{Tashiro_12,Boyarsky_12a,Boyarsky_12}. In Ref.\cite{Joyce_97}
(see also \cite{Cornwall_97}) it was argued that there can be more right-handed electrons 
over left-handed positrons due to some process in the early Universe at temperatures $T$ very much
higher than the electroweak phase transition(EWPT) scale ($\sim$100GeV). Their number is effectively conserved
at the energy scales much above the electroweak phase transitions and this allows one to introduce the chiral
chemical potentials $\mu_R(\mu_L)$. At temperature lower than $T_R\sim 80$~TeV  processes related
with the electron chirality flipping  may dominate over the Hubble expansion rate and the chiral chemical
potentials are not defined\cite{Campbell:1992jd,PhysRevD.49.6394,PhysRevLett.71.2372}. Further the right
handed current is not conserved due to the Abelian anomaly in the standard model (SM) and it satisfy the 
following equation:
\small{\begin{equation}
\partial_{\mu} \mathcal{J}^{\mu}_R=
-\frac{g^{\prime 2}y_{R}^2}{64 \pi^2}\mathcal{Y}^{\mu\nu} \tilde{\mathcal{Y}}^{\mu\nu}
=-\frac{g^{\prime 2}}{4\pi^2}\boldsymbol{\mathcal{E}}_Y\cdot \boldsymbol{B}_Y
\label{anomaly}
\end{equation}}
here, $\mathcal{Y}^{\mu\nu}=\partial^{\mu}Y^{\nu}-\partial^{\nu}Y^{\mu}$ is the field tensor associated with
the hyper-charge gauge field $Y^\mu$ and $\tilde{\mathcal{Y}}^{\mu\nu}=1/2\epsilon^{\mu\nu\rho\lambda}
\mathcal{Y}_{\rho\lambda}$. $\boldsymbol{\mathcal{E}}_Y$ and $\boldsymbol{\mathcal{B}}_Y$ respectively denote
hyper-electric and hyper-magnetic fields. Further, $g^{\prime}$ is associate gauge coupling, and $y_{R}$=-2
represents hypercharge of the right-electrons. The right-hand side of the first and second equality signs 
are related with the Chern-Simon number $N_{CS}$:
\begin{equation}
        N_{CS}=-\frac{g^{\prime 2}}{32\pi^2}\int d^3x\boldsymbol{\mathcal{B}}_Y\cdot \boldsymbol{\mathcal{Y}}
\label{CS}
\end{equation}
The anomaly equation (\ref{anomaly}) relates change in the right handed electron density with the variation 
of the topological (Chern-Simon or helicity) charge of the gauge fields. It has been shown in Ref.\cite{Semikoz:2011tm},
that CS term contributes in the effective standard model Lagrangian of the field $Y_{\mu}$, by polarization effect
through non-zero mean pseudo-vector current $\mathcal{J}_{j5}=g^{\prime 2}y_{R}^2/2<\bar{e}_R\gamma_j\gamma_5 e_R>
=-g^{\prime 2}y_{R}^2\mu_{eR} \mathcal{B}_j/4\pi^2$ and the effective Lagrangian for gauge field $Y_{\mu}$ in SM is 
\cite{Laine:2005bt,Semikoz:2011tm}.
\small{\begin{eqnarray}
 L_Y &=&-\frac{1}{4}\mathcal{Y}_{\mu\nu}\mathcal{Y}^{\mu\nu}-\mathcal{J}_{\mu}Y^{\mu}-\frac{g^{\prime 2}y_{R}^2\mu_{eR}}{4\pi^2}
\boldsymbol{\mathcal{B}}_Y\cdot \boldsymbol{Y}
\end{eqnarray}}
If these hyper-magnetic fields survives at the time of EWPT, they will give ordinary magnetic fields due to the electroweak mixing
$\mathcal{A_{\mu}}=cos \theta_w Y_{\mu}$. Where $Y_{\mu}$ is the
massless $U(1)_Y$ Abelian gauge hypercharge field.
It was shown that the chiral imbalance in the early Universe
can give rise to a magnetic field $B\sim 10^{22}$~G at temperature $T\sim$~100GeV with a typical
inhomogeneity scale $\sim 10^{6}/T$\cite{Joyce_97}.
In this work the authors have studied the Maxwell equations
with the Chern-Simon term and a kinetic equation consistent with equation (\ref{anomaly}). It was found that the transverse
modes can become unstable and give rise to the hypercharge magnetic field \cite{Son_12a,Avdoshkin:2014gpa}.
In Ref.\cite{Tashiro_12} the authors have used magnetohydrodynamics in the presence of chiral asymmetry
to study the evolution of magnetic field. They have shown that the chiral-magnetic\cite{Son_12a,Fukushima_08,Manuel:2015zpa}
and chiral-vorticity effects \cite{PhysRevD.20.1807} can play a significant role in the generation and dynamics of primordial magnetic field.
Further, it was demonstrated in Ref.\cite{Boyarsky_12} that evolution of the primordial magnetic field is strongly
influenced by the chiral anomaly even at a temperature as low as 10~MeV. It was shown that an isotropic and translationally
invariant initial state of the standard model plasma in thermal equilibrium can become unstable in the presence of
the global charges\cite{Boyarsky_12a}. The most general form of the polarization operator $\Pi^{ij}$ can be written
as:
\begin{equation}
        \Pi^{ij}(\vec{k})=\left(k^2\delta^{ij}-k^ik^j\right)\Pi_1(k^2)+i\epsilon^{ijk}k^k\Pi_2(k^2)
\label{setenor1}
\end{equation}
where, $\vec{k}$ is a wave-vector and $k^2=|\vec{k}|^2$. This equation satisfy the transversality
condition $k_i\Pi^{ij}=0$. It should be noted here that the Chern-Simon term is $\propto \boldsymbol{Y\cdot\partial Y}$, 
whereas the kinetic term is $\propto (\boldsymbol{\partial Y})^2$ and therefore the Chern-Simon term can dominate
over the large length scales.
Thus, a non-zero value of $\Pi_2$ when $k\rightarrow 0$ implying the presence
of Chern-Simon term in the expression for the free energy. Using the field theoretic framework
in \cite{Boyarsky_12a} it was shown that for a sufficiently small $k < \Pi_2(k^2)/\Pi_1(k^2)$, the polarization tensor $\Pi^{ij}$ has a negative
eigenvalue and the corresponding eigen mode gives the instability.

 Recently there has been an interesting development in incorporating the parity-violating effects
into a kinetic theory formalism (see Ref.\cite{Son_12a,Stephanov_12,Son_13,Chen_13}). In this
approach the kinetic (Vlasov) equation is modified by including the Berry curvature term which
takes into account chirality of the particles. The modified kinetic equation is consistent with
the anomaly equation (\ref{anomaly}). Incorporation of the parity odd physics in kinetic theory
leads to a redefinition of the Poisson brackets which
includes contribution from the Berry connection. The confidence that the new kinetic equation
captures the proper physics stems from the fact that the equation is consistent
with the anomaly equation (\ref{anomaly}) and it also reproduces some of the known results obtained
using the quantum field theory with the parity odd interaction\cite{Akamatsu_13}.
In fact the \textquotedblleft classical" kinetic equation
can reproduce, in the leading order in the hard dense loop approximation, the parity-odd correlation
of the underlying quantum field theory \cite{Laine:2005bt,PhysRevLett.54.970}.
The modified kinetic equation can also be derived from
the Dirac Hamiltonian by performing semiclassical Foldy-Wouthuysen diagonalization \cite{Manuel_14,Silenko:2007wi}.
The modified kinetic equation can be applied to both the high density or high temperature regimes\cite{Manuel_14}.
Further in Ref.\cite{Akamatsu_13} normal modes of the chiral plasma were analyzed
using the modified kinetic theory in the context of heavy-ion collisions.
In this work authors have found that
in the collisionless limit transverse branch of the dispersion relation
can become unstable with typical wave number $k\sim \alpha^{\prime}\mu/\pi$, where
$\alpha^{\prime}$ is the coupling constant and $\mu$ refers to the chiral chemical potential.

Here we would like to note that the authors in Ref.\cite{Joyce_97} have
used a heuristically written kinetic equation which is consistent with the equation (\ref{anomaly})
to study generation of the primordial magnetic field. In addition authors have
used expression for the current by incorporating standard electric resistivity
and the chiral-magnetic effect. The chiral-vorticity effect was not considered.
It should be emphasized that  the forms of the kinetic equation used in Ref.\cite{Joyce_97}
and the Berry curvature modified theory\cite{Son_13} are very different. As both
the approaches describe the same physics, it would be interesting to see under what conditions they give
the similar predictions. Keeping the above discussion in mind we believe that it would be highly useful to
consider the problem of generation of primordial magnetic field in presence
of the Abelian anomaly by using the Berry curvature modified kinetic theory.
In this work we incorporate the effect of collisions in the modified kinetic
theory and derive expressions for the electric and magnetic resistivities. The new kinetic framework also allows us to calculate the generation of the primordial magnetic field
and vorticity. Further in our calculation, we considered an isotropic and homogeneous
initial state of the particle distribution function.
The magnetic field
is generated by the unstable transverse modes in presence of chiral charges ($Q_5$). This can be seen by integrating equation (\ref{anomaly}) over space. One can write: $\partial_0\left(Q_5+ \frac{\alpha^{\prime}}{\pi}\mathcal{H} \right)=0$, where,
$Q_5=\int j^0d^3x$ and the helicity $ \mathcal{H} =\frac{1}{V} \int d^3 x(\boldsymbol{Y}\cdot \boldsymbol{\mathcal{B}_Y})$. The finite helicity state can be
created even if the initial state has $\mathcal{H}=0$ but $Q_5\neq0$. Thus the magnetogenesis by net non-zero
chiral-charges may not require any pre-existing seed field.

The manuscript is organized into three sections. In section-I we briefly state the (3+1) formalism
of Thorne and Macadonald \cite{Macadonald_82} and the kinetic theory with the Berry curvature.
In section-II we apply this formalism to study the primordial magnetic field generation
in presence of the chiral asymmetry. We also calculated the vorticity generation in the plasma due to the chiral imbalance.
Section-III, contains results and  a brief discussion. We have shown that our estimated value of the peak magnetic field actually falls within the constraints obtained current observations.
\section{Basic Formalism }
\subsection*{Maxwell's equations in the expanding universe}
 In this work we shall study the generation of primordial magnetic field at the time when temperature
of the Universe was much higher than $T_R\sim80$~TeV(much higher than the EWPT temperature).
We intend to solve the coupled system of the modified kinetic and the Maxwell's equations in the expanding
universe background. Here we note that we ignore the fluctuations in the metric
due to the matter perturbation. For this one needs to write the underlying equations in a general covariant
form. Interestingly the techniques developed in Refs.\cite{Dettmann_93,PhysRevD.40.3809,Gailis_94,Gailis_95} allows
one to write the system of kinetic and Maxwell's
equations in the expanding background into the form that look similar to their flat space-time form.
In this formalism the well developed intuition and techniques of the flat space-time plasma physics can be exploited
to study the problem at hand.
This can be accomplished by choosing a particular set of fiducial observers (FIDO's)\cite{Dettmann_93} at each point
of space-time with respect to which all the physical quantities including hyper-electric and magnetic
fields are measured. A line element for the expanding background can be written using the
Friedmann-Lema\^{\i}tre-Robertson-Walker metric as
\begin{equation}
 ds^{2}=- dt^2+a^2(t)(dx^2+dy^2+dz^2)
\end{equation}
where  $x$, $y$ and $z$ represent comoving coordinates. Here t is the proper time seen by observers
at  the fixed $x$, $y$ \& $z$ and a(t) is scale factor.
One can introduce the conformal time $\eta$ using the definition
 $\eta =\int dt/a^2(t)$ to write this metric as:
\begin{equation}
 ds^{2}= a^2(\eta)(-d\eta^2+dx^2+dy^2+dz^2))
\end{equation}
The (hyper)-electric $\boldsymbol{E}_{phy}$, (hyper)-magnetic $\boldsymbol{B}_{phy}$ and the
current density $\boldsymbol{J}_{phy}$ are related to the corresponding fiducial quantities by transformations:
$\boldsymbol{\mathcal{E}}=a^2\boldsymbol{E}_{phy}$, $\boldsymbol{\mathcal{B}}=
a^2\boldsymbol{B}_{phy}$,
$\boldsymbol{\mathcal{J}}=a^3 \boldsymbol{J}_{phy}$.
One can now write the Maxwell's equations in the fiducial frame as:
\small{\begin{gather}
 \frac{\partial {\boldsymbol{\mathcal{B}}}}{\partial \eta}+\boldsymbol{\nabla}\times\boldsymbol{\mathcal{E}}=0 \label{M1}\\
 {\boldsymbol{{\nabla}\cdot\mathcal{E}}}=4\pi\rho_e \label{M2} \\
 {\boldsymbol{{\nabla}\cdot{\mathcal{B}}}}=0 \label{M3}\\
 {\boldsymbol{{\nabla}\times{\mathcal{B}}}}= 4\pi {\boldsymbol{\mathcal{J}}}+\frac{\partial {{\boldsymbol{\mathcal{E}}}}}{\partial \eta} \label{M4}
\end{gather}}
where $\boldsymbol{\mathcal{B}}$, $\boldsymbol{\mathcal{E}}$, $\rho_e$ and $\boldsymbol{\mathcal{J}}$ are respectively magnetic field, electric field, charge density and current density seen by the fiducial observer.
\subsection*{Kinetic theory with Berry curvature}
 The charge $\rho_e$ and current $\boldsymbol{\mathcal{J}}$ in the Maxwell's equations (\ref{M1}-\ref{M4}) can be calculated
 using the Berry curvature modified kinetic equation which is also consistent with the quantum anomaly equation.
The modified kinetic equation is given by
\small{\begin{align}\label{evolution equation}
\frac{\partial {f}}{\partial \eta}+&
 \frac{1}{1+e\boldsymbol{\Omega_{\boldsymbol{p}}}\cdot\boldsymbol{\mathcal{B}}}
 \bigg[(e\boldsymbol{\Tilde{\mathcal{E}}}+e{\boldsymbol{\Tilde{v}}\times\boldsymbol{\mathcal{B}}}
 +e^2{(\boldsymbol{\Tilde{\mathcal{E}}}\cdot{\mathcal{B}}) \boldsymbol{\Omega_p})}\cdot\frac{\partial{f}}{\partial{\boldsymbol{p}}} \nonumber \\
 &+({\boldsymbol{\Tilde{v}}}+e {\boldsymbol{\Tilde{E}}\times\boldsymbol{\Omega_p}}+e({\boldsymbol{\Tilde{v}}\cdot\boldsymbol{\Omega_p})
 \boldsymbol{\mathcal{B}}})\cdot
 \frac{\partial{f}}{\partial{\boldsymbol{r}}}\bigg]=\left(\frac{\partial f}{\partial \eta}\right)_{coll}
\end{align}}
where $e$ is charge of the particles and has relation with the electroweak(EW) mixing angle $\theta_w$: $e= g^{\prime}cos \theta_w$.
Also the $g^{\prime}$ is related with the $U(1)_Y$ gauge coupling constant $\alpha^{\prime}$ as $\alpha^{\prime}=g^{\prime 2}
cos^2 \theta_w/4\pi$. $\boldsymbol{\Tilde{v}}=\partial \epsilon_{\boldsymbol{p}}/\partial{\boldsymbol{p}}=\boldsymbol{v}$
and $e \boldsymbol{\Tilde{\mathcal{E}}}= e \boldsymbol{\mathcal{E}}-\partial{\epsilon_{\boldsymbol{p}}}/\partial
\boldsymbol{r}$. $\boldsymbol{\Omega_{p}}= \pm \boldsymbol{p}/(2p^3)$ is Berry curvature. $\epsilon_{\boldsymbol{p}}$
is defined as $\epsilon_{\boldsymbol{p}}=p(1-e\boldsymbol{\mathcal{B}}
\cdot\boldsymbol{\Omega_{p}})$ with $p=|\boldsymbol{p}|$. The positive sign corresponds to right-handed
fermions where as the negative sign is for left-handed ones. In the absence of Berry correction i.e.
$\boldsymbol{\Omega_p}=0$, above equation  reduces to the Vlasov equation when the collision term on the right
hand side of equation (\ref{evolution equation}) is absent.

 As we have already stated that we are interested in temperature regime $T>T_R\gg T_{EW}$. At these
temperatures electrons are massless. The only processes that
can change electron chirality is its Yukawa interaction with Higgs
But at this temperature this interaction is not strong enough to alter electron chirality.
It is here important to note that for temperature
smaller than $T_R$ electron mass plays major role in left-right asymmetry. Recently in Ref.\cite{PhysRevD.91.085035}
it has been shown that at temperature of the order of MeV, the mass of the electron plays an important
role in determining the magnetic properties of the
proto-neutron star by suppressing the chiral charge density during the core collapse of supernova.
However for the present case we ignore the electron mass
by considering only $T>T_R$ regime. Thus we write particle
number density modified by the Berry term as:
\small{\begin{equation}
 N=\int \frac{d^3p}{(2\pi)^3}(1+e{\boldsymbol{\mathcal{B}.\Omega_p}})f
\end{equation}}
Above equation (\ref{evolution equation}) can be converted to following form by multiplying by $(1+e\boldsymbol{\mathcal{B}}\cdot\boldsymbol{\Omega_p})$ 
and integrating over p
\small{\begin{equation} \label{conservation}
 \frac{\partial N}{\partial{\eta}}+{\boldsymbol{\nabla\cdot\mathcal{J}}} = - e^2\int \frac{d^3p}{(2\pi)^3}\bigg({\boldsymbol{\Omega_p}}
 \cdot\frac{\partial f}{\partial \boldsymbol{p}}\bigg) (\boldsymbol{\mathcal{E}\cdot\mathcal{B}}),
\end{equation}}
In equation (\ref{conservation}) $\boldsymbol{\mathcal{J}}$ is total current and is defined as
$\boldsymbol{\mathcal{J}}= \Sigma_{a}\boldsymbol{\mathcal{J}}_a$. Here index $a$ denotes
current contribution from different species of the fermion e.g. right-left particles and there antiparticles. $\boldsymbol{\mathcal{J}}_a$ is defined
as:
\small{\begin{align} \label{current1}
 \boldsymbol{\mathcal{J}}_{a}= - e^a\int & \frac{d^3p}{(2\pi)^3}\bigg[\epsilon_{\boldsymbol{p}}^a\frac{\partial {f}_{a}}{\partial \boldsymbol{p}}
 +e^a(\boldsymbol{\Omega_{\boldsymbol{p}}^a}\cdot\frac{\partial{f}_{a}}{\partial{\boldsymbol{p}}})\epsilon_{\boldsymbol{p}}^a\mathcal{B} \nonumber \\
 &+\epsilon_{\boldsymbol{p}}\boldsymbol{\Omega_{\boldsymbol{p}}^a}\times\frac{\partial{f}_{a}}{\partial \boldsymbol{r}}\bigg]
 +e^a({\boldsymbol{\mathcal{E}}\times\boldsymbol{\sigma}}^a)
\end{align}}
where $\sigma^a =\smallint d^3p/(2\pi)^3 \boldsymbol{\Omega_p}^a f_a$ and $\epsilon_{\boldsymbol{p}}^a=p(1-e^a\boldsymbol{\mathcal{B}}
\cdot\boldsymbol{\Omega_{p}}^a)$ with $p=|\boldsymbol{p}|$.
Depending on the species, charge $e$, energy of the particles $\epsilon_p$, Berry curvature $\boldsymbol{\Omega_p}$
and form of the distribution function $f$ changes. For the right-handed particle $a=R$ and hyper-charge $e$ and for right-handed
anti-particle $a=\bar{R}$ and for charge this case is -$e$ etc. 
It is clear from equation (\ref{conservation})
that in presence of external electric and magnetic fields the chiral current no longer conserved. The first term in equation (\ref{current1})
is usual current equivalent to the kinetic theory and
 remaining second and  third terms are the current contribution by Berry correction. The last term is due to the
 anomalous Hall effect and  it  vanishes for a spherically symmetric distribution function.
 If we follow the power counting scheme as in \cite{Son_12a} i.e. $\boldsymbol{Y_{\mu}}=\mathcal{O}(\epsilon)$ ,
 $\partial_r=\mathcal{O}(\delta)$ (where $\boldsymbol{Y_{\mu}}$ represents $U(1)_Y$
gauge field) and considering only terms of the order of $\mathcal{O}(\epsilon\delta)$ in equation (\ref{evolution equation}), we will have:
\small{\begin{equation}
 \bigg(\frac{\partial }{\partial\eta}+{\boldsymbol{v}}\cdot\frac{\partial}{\partial \boldsymbol{r}}\bigg)f_a+
 \bigg(e^{a}\boldsymbol{\mathcal{E}}+e^{a}(\boldsymbol{v}\times \boldsymbol{\mathcal{B}})- 
 \frac{\partial{\epsilon^{a}_p}}{\partial{\boldsymbol{r}}}\bigg)\cdot\frac{\partial f_a}{\partial \boldsymbol{p}}=
 \bigg(\frac{\partial f_a}{\partial\eta}\bigg)_{coll.}
 \label{vlasov1}
\end{equation}}
where, we have taken $\boldsymbol{v}=\boldsymbol{p}/p$.
In the subsequent discussion we shall apply this equation to study the evolution of the primordial magnetic field.
\subsection*{Current and polarization tensor for chiral plasma}
 Here we assume that the plasma of the standard particles is in a state of \textquoteleft thermal-equilibrium\textquoteright ~at temperature $T>T_R$
 and at these temperatures the masses of the plasma particles can be ignored. We also assume that there exist a left-right
 asymmetry and there is no large-scale electromagnetic field. Thus the equilibrium plasma considered to be in a homogeneous
 and isotropic state which is similar to the assumptions made in Ref.\cite{Joyce_97,Boyarsky_12a}.
For a homogeneous and
isotropic conducting plasma in thermal equilibrium, distribution function for different species are:
\begin{equation}
 f_{0a}(p)=\frac{1}{exp(\frac{\epsilon_p^0-\mu_a}{T})+1}
\end{equation}
If $\delta f_R$ and $\delta f_{\bar{R}}$ are fluctuations in the distribution functions of the right electron and right-antiparticles around
there equilibrium distribution. Then we can write perturbed distribution functions as
\begin{eqnarray}
 f_{R}({\boldsymbol{r},\boldsymbol{p}}, \eta)= f_{0R}(p)+ \delta f_{R}({\boldsymbol{r, p}}, \eta) \\
 f_{\bar{R}}({\boldsymbol{r,p}},\eta)=f_{0\bar{R}}(p)+ \delta f_{\bar{R}}({\boldsymbol{r,p}},\eta)
\end{eqnarray}
Subtracting equation for $a=\bar{R}$ from $a=R$ using equation (\ref{vlasov1}) one can write:
\small{\begin{align}
 \bigg(\frac{\partial}{\partial\eta}+ {\boldsymbol{v}}\cdot\frac{\partial}{\partial {\boldsymbol{r}}}\bigg)f({\boldsymbol{r,p}},\eta)+&
 ep\frac{\partial(\boldsymbol{\mathcal{B}\cdot \Omega_p})}{\partial\boldsymbol{r}}\cdot 
 \frac{\partial f_0}{\partial \boldsymbol{p}} 
 + ({\boldsymbol{\mathcal{E}.v}})\frac{d {f_0}}{dp} \nonumber \\
 &=\left(\frac{\partial f(\boldsymbol{r,p},\eta)}{\partial\eta}\right)_{coll.} \label{R-R_}
\end{align}}
where, $f(\boldsymbol{r,p},\eta)=(f_R-f_{\bar{R}})$ and $f_{0}=f_{0R}+f_{0\bar{R}}$.
Here we have used $\frac{\partial f^0}{\partial\boldsymbol{p}}=\boldsymbol{v}\frac{df^0}{dp}$.
This equation relates the fluctuations of the distribution functions of the charged particles with the induced gauge field fluctuations.
The gauge field fluctuations can be seen from the Maxwell's electromagnetic equations (\ref{M1}-\ref{M4}).
Under the relaxation time approximation, the collision term can be written as $({\partial f_a}/{\partial\eta})_{coll.}\approx -\nu_{c}(f_a-f_{0a})$
(one can also look at some studies in chiral kinetic theory with collision in Ref. \cite{Chen:2014cla,Chen:2015gta}.
Next, we take Fourier transform of the all perturbed quantities namely $\boldsymbol{\mathcal{E}}$, $\boldsymbol{\mathcal{B}}$ and
$f(\boldsymbol{r,p},\eta)$ by considering the spatio-temporal variation of these quantities as
$ exp[-i(\omega\eta-\boldsymbol{k\cdot r})]$.
Then using equation (\ref{R-R_}) one can get
\small{\begin{equation}
 f_{{\boldsymbol{k}},\omega}=\frac{-e[({{\boldsymbol{v}\cdot \boldsymbol{\mathcal{E}}_{\boldsymbol{k}}}})
 +\frac{i}{2p}({\boldsymbol{v}\cdot\boldsymbol{\mathcal{B}}_{\boldsymbol{k}}})(\boldsymbol{k\cdot v})]
 \frac{df_0}{dp}}{i({\boldsymbol{k\cdot v}}  -\omega-i\nu_c)}\label{f_{kw}}.
\end{equation}}
So current contribution for the right handed particle and right handed antiparticles in terms of mode 
function can be written using equation (\ref{current1}) as (ignored anomalous hall current part):
\small{\begin{align} 
 \boldsymbol{\mathcal{J}}_{\boldsymbol{k}R}= & e\int \frac{d^3p}{(2\pi)^3}\bigg[\{{\boldsymbol{v}}-\frac{i}{2p}\boldsymbol{(v\times k)}\}
 f_{\boldsymbol{k}\omega R} \nonumber \\
 &-\frac{e}{2p^2} \{\boldsymbol{\mathcal{B}_k} 
 -\boldsymbol{v}(\boldsymbol{v}\cdot \boldsymbol{\mathcal{B}}_{\boldsymbol{k}})\}f_{0}
 +\frac{e}{2p}\boldsymbol{\mathcal{B}}_{\boldsymbol{k}}\frac{df_{0}}{dp}\bigg]\label{J_{kw}} 
\end{align}}
%
In the similar way we can get current contribution from left handed particle and left handed antiparticle. 
So we can obtain total current by adding the contributions from both left and right handed particles and 
antiparticles by putting perturbations $f_{k\omega}$ for all species in equation (\ref{J_{kw}}) and adding as:
\small{\begin{align} \label{j total}
 \boldsymbol{\mathcal{J}}_{\boldsymbol{k}} = &-m_D^2 \int \frac{d\Omega}{4\pi} \frac{\boldsymbol{v}(\boldsymbol{v}.
 \mathcal{\boldsymbol{E_{k}}})}{i(\boldsymbol{k.v}-\omega-i \nu_c)} \nonumber \\
  &-\frac{h_D^2}{2} \int \frac{d\Omega}{4\pi}\{\boldsymbol{\mathcal{B}}_k-\boldsymbol{v}(\boldsymbol{v. \mathcal{B}_k})\}\nonumber \\
 &-\frac{i g_D^2}{4}\int \frac{d\Omega}{4\pi} \frac{(\boldsymbol{v \times k})(\boldsymbol{v.\mathcal{B}_k})(\boldsymbol{k.v})}{
 (\boldsymbol{k.v}-\omega-i \nu_c)}\nonumber \\
 &-\frac{c_D^2}{2}\int\frac{d \Omega}{4\pi} \bigg\{\frac{\boldsymbol{v} (\boldsymbol{v.\mathcal{B}_{k}})(\boldsymbol{k.v})-
 (\boldsymbol{v\times k})(\boldsymbol{v.\mathcal{E}_{k}})}{(\boldsymbol{k.v}-\omega-i \nu_c)}+ \boldsymbol{\mathcal{B}}_{k}\bigg\}
\end{align}}
Here $\Omega$ represent angular integrals. In equation (\ref{j total}), we have defined $m_D^2 =e^2\int \frac{p^2 dp}{2\pi^2}
\frac{d}{dp}( f_{0R}+f_{0\bar{R}}+f_{0L}+f_{0\bar{L}})$,
$c_D^2 =e^2\int \frac{p dp}{2\pi^2}\frac{d}{dp}( f_{0R}-f_{0\bar{R}}-f_{0L}+f_{0\bar{L}})$, $g^2_D=e^2\int
\frac{dp}{2\pi^2}\frac{d}{dp}( f_{0R}+f_{0\bar{R}}+f_{0L}+f_{0\bar{L}})$, $h^2_D=e^2\int \frac{dp}{2\pi^2}
( f_{0R}-f_{0\bar{R}}-f_{0L}+f_{0\bar{L}})$.

      Expression for the polarization tensor $\Pi^{ij}$ can be obtained from equation (\ref{j total}) by writing total 
      current in the following form
      $\mathcal{J}^{i}_{k}=\Pi^{ij}(k)Y_{j}(k)$ using $\boldsymbol{\mathcal{E}_{k}}=-i \omega \boldsymbol{Y_{k}}$
 and $\boldsymbol{\mathcal{B}_{k}}= i (\boldsymbol{k}\times \boldsymbol{Y_{k}})$.
 One can express $\Pi^{ij}$ in terms of longitudinal $P^{ij}_{L}= k^{i}k^{j}/k^2$, transverse  $P^{ij}_{T}= (\delta^{ij}-k^{i}k^{j}/k^2)$
 and the axial $P^{ij}_{A}=i\epsilon^{ijk}k^k$ projection operators as $\Pi^{ik}=\Pi_{L}P_{L}^{ik}+\Pi_{T}P_{T}^{ik}+\Pi_{A}P_{A}^{ik}$.
 After performing the angular integrations in equation (\ref{j total}) one obtains $\Pi_L$, $\Pi_T$ and $\Pi_A$ as given below
\small{\begin{gather} \label{polarization}
\Pi_{L} =-m_D^2\frac{\omega\omega^{\prime}}{k^2}[1-\omega^{\prime} L(k)],\dotfill\\
\Pi_{T} = m_D^2\frac{\omega\omega^{\prime}}{k^2}\bigg[1+\frac{k^2-\omega^{\prime 2}}{\omega^{\prime}}L(k)\bigg],\dotfill\\
\Pi_{A} =-\frac{h_{D}^2}{2}  \bigg[1-\omega(1-\frac{\omega^{\prime 2}}{k^2})L(k)-\frac{\omega^{\prime}\omega}{k^2}\bigg]
\end{gather}}
where, ${\textstyle L(k)=\frac{1}{2k}ln\bigg(\frac{\omega^{\prime}+k}{\omega^{\prime}-k}\bigg)}$ and ${\textstyle \omega^{\prime}= 
\omega+i \nu_{c}}$. Also ${\textstyle m^2_D= 4\pi \alpha^{\prime} \left(\frac{T^2}{3}+\frac{\mu^2_R+\mu^2_L}{2\pi^2}\right)}$
and ${\textstyle h^2_D= \frac{2\alpha^{\prime}\Delta\mu}{\pi}}$. We have defined ${\textstyle \Delta\mu=(\mu_R-\mu_L)}$.
In above integrals, we have replaced, $e$ by $\alpha^{\prime}$
using relation $e^2=4\pi \alpha^{\prime}$ . First,  consider case when  $\nu_c=0$.
In the limit $\omega\rightarrow 0$, $\Pi_L$ and $\Pi_T$  vanish and
the parity odd part of the polarization tensor $\Pi_A= h_D^2/2\approx \alpha^{\prime} \Delta\mu/\pi $. Here it should be noted that
   $\Pi_A$ does not get thermal correction. This could be due to the fact that origin
of $\Pi_A$ term is related with the axial anomaly and it is well known that anomaly does not receive any thermal
correction \cite{PhysRevD.38.3840,Itoyama:1982up,TGomezNicola:1994vq}. This form of $\Pi_A$ is similar to the result
obtained in \cite{Boyarsky_12a} using quantum field theoretic arguments at $T\leq 40$~GeV. But in the kinetic
theory approach presented here such no assumption is made.  Normal modes for the plasma can be obtained by
using expressions for $\Pi_L$, $\Pi_T$ and $\Pi_A$. Using equation $\partial _{\nu}F^{\mu\nu}= -4\pi\mathcal{J}^{\mu}$, we
can write to the following relation
\begin{equation}
[M^{-1}]^{ij}Y_{j}(k)=-4\pi\mathcal{J}^{i}_{k},
\end{equation}
where, $[M^{-1}]^{ik}=[(k^2-\omega^2)\delta^{ik}-k^ik^k+\Pi^{ik}]$.
Dispersion relations can be obtained from the poles of $[M^{-1}]^{ik}$, which are as given below
\begin{gather*}
 \omega^2=\Pi_L,\\
 \omega^2= k^2+\Pi_{T}(k)\pm k\Pi_{A}.
\end{gather*}
One can study the normal modes of the chiral plasma and instabilities using these dispersion relations. However,
it is more instructive to study dynamical evolution of magnetic field by explicitly writing
time dependent Maxwell equations.
\section{Generation of the primordial magnetic field and vorticity}
   Plasmas with chirality imbalance are known to have instabilities that can generate magnetic fields
in two different regimes: (i) collision dominated $k, \omega\ll\nu_c$ \cite{Joyce_97} and
(ii) collisionless case i.e. $\nu_c=0$ \cite{Akamatsu_13}.
In this section we analyze how the magnetic fields evolve in the plasma due to these instabilities,
within the modified kinetic theory frame work.  Expression for the total current described by Eq.(\ref{j total})
can be written as $\mathcal{J}_{k}^{i}= \sigma_{E}^{ij}{\mathcal{E}_k}^{j}+
\sigma_{B}^{ij}\mathcal{B}_{k}^{j}$ where $\sigma_{E}^{ij}$ and
$\sigma_{B}^{ij}$ are electrical and magnetic conductivities.
The integrals involved in equation (\ref{j total}) are rather easy to evaluate in the limit $k,\omega \ll \nu_c$ and one can write
the expression for $\sigma_E^{ij}$ and $\sigma_B^{ij}$  as:
\small{\begin{gather} 
 \sigma_{E}^{ij} \approx 
 \left(\frac{m_D^2}{3\nu_c}\delta^{ij}+\frac{i}{3\nu_c}\frac{\alpha^{\prime}
 \Delta \mu}{\pi}\epsilon^{ijl}k^{l}\right)    \label{sigma E}\\
 \sigma_{B}^{ij} \approx
 -\frac{4}{3}\frac{\alpha^{\prime} \Delta \mu}{\pi} \delta^{ij}  \label{sigma B}   
\end{gather}}
Here, we would like to note that the Berry curvature correction in the kinetic equation gives us an additional contribution
in the expression for $\sigma^{ij}_E$ which was not incorporated in Ref.\cite{Joyce_97}.
First term is the usual dissipative part of the electric current
and it contributes to the Joules dissipation. The second term
is due to the chiral imbalance and it does not give any contribution
to the Joules heating.
 As we shall demonstrate
later this term is responsible for the vorticity current \cite{Fukushima_08}.
 One can write the total current as $\mathcal{J}^{i}_{\boldsymbol{k}}=\sigma^{ij}_E\mathcal{E}^{j}_{\boldsymbol{k}}
 +\sigma^{ij}_B \mathcal{B}^{j}_{\boldsymbol{k}}$ and the Maxwell's equation:
$i(\boldsymbol{k\times \mathcal{B}_k})^{i}= 4\pi\mathcal{J}^{i}_{\boldsymbol{k}}$. Here we have
dropped the displacement current term (this is valid when $\frac{\omega}{4\pi\sigma}\ll 1$).
Next by taking vector product of $\boldsymbol{k}$ with the above Maxwell equation
one obtains (using Maxwell's equations and after some simplification):
\small{\begin{align} \label{B_kw vector}
 \frac{\partial \boldsymbol{\mathcal{B}}_k}{\partial \eta}+\left(\frac{3\nu_c}{4\pi m_d^2}\right)k^2\boldsymbol{\mathcal{B}}_k 
& -\left(\frac{\alpha^{\prime}\Delta\mu}{\pi m_D^2}\right)\left(\boldsymbol{k}\times
(\boldsymbol{k}\times\boldsymbol{\mathcal{E}_k})\right) \nonumber \\
&+i\frac{4\alpha^{\prime} \nu_c \Delta\mu}{\pi m_D^2} (\boldsymbol{k\times\mathcal{B}_k})=0.
\end{align}}
This is the magnetic diffusivity equation for the chiral plasma. By replacing $(\boldsymbol{k \times \mathcal{E}_k})$ by
$-\frac{1}{i}\frac{\partial \boldsymbol{\mathcal{B}_k}}{\partial \eta}$
in equation (\ref{B_kw vector}), we can solve this equation without a loss of generality by considering
the propagation vector $\boldsymbol{k}$ in $z-$direction and the magnetic field having components perpendicular to $z-$ axis.
After defining two new variables:
$\tilde{\mathcal{B}_k}= (\mathcal{B}_k^1+ i\mathcal{B}_k^2)$ and
$\tilde{\mathcal{B}^{\prime}_k}= (\mathcal{B}_k^1- i\mathcal{B}_k^2)$ one
can rewrite equation (\ref{B_kw vector}) as,
\small{\begin{eqnarray}
 \frac{\partial\tilde{\mathcal{B}}_{k}}{\partial \eta}+\left[\frac{\left(\frac{3\nu_c}{4\pi m_d^2}\right)k^2-
 \left(\frac{4\alpha^{\prime}\nu_c \Delta\mu}{\pi m_D^2}\right)k}
 {(1+\frac{\alpha^{\prime}\Delta \mu k}{\pi m_D^2})}\right]\tilde{\mathcal{B}}_{k}=0, \label{mode1} \\
  \frac{\partial\tilde{\mathcal{B}}_{k}^{\prime}}{\partial \eta}+\left[\frac{\left(\frac{3\nu_c}{4\pi m_d^2}\right)k^2+
  \left(\frac{4\alpha^{\prime} \nu_c \Delta\mu}{\pi m_D^2}\right)k} {(1-\frac{\alpha^{\prime}\Delta \mu k}{\pi m_D^2})}\right]
  \tilde{\mathcal{B}}_{k}^{\prime}=0.\label{mode2}
\end{eqnarray}}
 Thus the magnetic field vector $\mathcal{B}$
can be decomposed into these new variables $\tilde{\mathcal{B}_k}$ \& $\tilde{\mathcal{B}^{\prime}_k}$ having definite helicity
(or circular polarization). The effect of Ohmic decay
is already there into the above equations due to inclusion of collision frequency $\nu_c$.
It should be noted here that if $\alpha^{\prime}\Delta \mu k/\pi m_D^2 \ll 1$ equation (\ref{mode1}) is similar
to the magnetic field evolution equation considered in Ref.\cite{Joyce_97}. In this limit equation (\ref{mode2})
will give a purely damping mode. In this case dispersion relation will be 
\begin{equation} \label{dispersr}
 i \omega =\frac{3\nu_c}{4\pi m_d^2}k^2- \frac{4\alpha^{\prime}\nu_c \Delta\mu}{\pi m_D^2}k
\end{equation}
In Appendix we have shown that the dispersion relation we have found here using kinetic theory
matches with the dispersion relation obtained in \cite{Akamatsu_13}. 

 The instability can also occur in the collisionless regime ($\nu_c=0$) \cite{Akamatsu_13}. In
the quasi-static limit i.e. $\omega\ll k$ one can define the electric conductivity
as $\sigma^{ij}_E\approx \pi(m^2_D/2k)\delta^{ij}$ and magnetic conductivity
$\sigma^{ij}_B\approx (h^2_D/2)\delta^{ij}$. Here it should be noted that the
above conductivities do not depend upon the collision frequency. Similar to the previous case one can take the propagation vector in
$z-$direction and consider components of the magnetic field in the direction perpendicular $z-$
 axis. One can write a set of decoupled equations describing the evolution of magnetic field using the
variables $\tilde{\mathcal{B}}_{k}$ and $\tilde{\mathcal{B}_k}^{\prime}$ as:
\small{\begin{eqnarray}
 \frac{\partial\tilde{\mathcal{B}}_{k}}{\partial \eta}+\Bigg[\frac{k^2-\frac{4\alpha^{\prime} \Delta \mu k}{3}}{\frac{\pi m_D^2}{2k}}
 \Bigg]\tilde{\mathcal{B}}_{k}=0, \label{mode11}\\
 \frac{\partial\tilde{\mathcal{B}}_{k}^{\prime}}{\partial \eta}+\Bigg[\frac{k^2+\frac{4\alpha^{\prime}
 \Delta \mu k}{3}}{\frac{\pi m_D^2}{2k}}
 \Bigg]\tilde{\mathcal{B}_k}^{\prime}=0.\label{mode22} 
\end{eqnarray}}
Here we note that if one replaces $\partial/\partial\eta$ by $-i\omega$ equations (\ref{mode11})
and (\ref{mode22}) gives the same dispersion relation for the instability as discussed in Ref.\cite{Akamatsu_13}.
\subsection*{Vorticity generated from chiral imbalance in the plasma}
It would be interesting to see if the instabilities arising due to chiral-imbalance can lead to
vorticity generation in the plasma. In order to study vorticity of the plasma, we define the average velocity as:
\small{\begin{equation} \label{vorticity}
 <\boldsymbol{v}> = \frac{1}{\bar n}\int \frac{d^3p}{(2\pi)^3}\boldsymbol{v} (\delta f_R- \delta f_{\bar{R}}+\delta f_L-
 \delta f_{\bar{L}})
\end{equation}}
Here we have used the perturbed distribution function in the numerator of the above equation
which is due to the fact that the equilibrium distribution function is assumed to homogeneous and
isotropic and therefore will not contribute to vorticity dynamics.
The denominator is total number density and is defined as (in Ref.\cite{turner} page number 63):
\begin{align}
\bar n =&n_{particle}-n_{antiparticle} \nonumber\\
= & d_f\int_0^{\infty} \frac{d^3p}{(2\pi)^3}\left(\frac{1}{1+exp(\frac{p-\mu}{T})}-\frac{1}{1+exp(\frac{p+\mu}{T})}\right)
\end{align}
which in the case of chiral plasma gives $\bar n= \frac{2}{3} T^2 (\mu_R +\mu_L)$. 
We consider $k, \omega\ll\nu_c$ regime, 
in this case the perturbed distribution function say for the right-handed particles can be written as:
\small{\begin{equation} 
 \delta f_{{\boldsymbol{k}},\omega R}=-\frac{e}{\nu_c}[({{\boldsymbol{v}\cdot \boldsymbol{\mathcal{E}}_{\boldsymbol{k}}}})
 +\frac{i}{2p}({\boldsymbol{v}\cdot\boldsymbol{\mathcal{B}}_{\boldsymbol{k}}})(\boldsymbol{k\cdot v})]\frac{df_{0R}}{dp} 
 \label{R perturbed}
\end{equation}}
If we add the contribution for all the particles species and their anti-particles one can write the numerator in equation (\ref{vorticity}) as:
\small{\begin{equation}
\sqrt{\frac{\alpha^{\prime}}{\pi^3}} \frac{1}{\nu_c}\left(\frac{T^2}{3}+\frac{\mu_R^2+\mu_L^2}{2\pi^2}\right)\boldsymbol{\mathcal{E}}_k
\approx \sqrt{\frac{\alpha^{\prime}}{\pi^3}} \frac{T^2}{3\nu_c}\boldsymbol{\mathcal{E}}_k
\end{equation}}
Above we have neglected $\frac{3}{2\pi^2}\frac{\mu_R^2+\mu_L^2}{T^2}$ in comparison to one, as the value of $\mu_R/T$ and $\mu_L/T$
are very small ($O(10^{-4})$). One can write average velocity as follows
\small{\begin{equation} \label{mu}
 <\boldsymbol{v}_k> =\sqrt{\frac{\alpha^\prime}{\pi^3}} \frac{1}{2\nu_c}\frac{1}{\mu_R+\mu_L}\boldsymbol{\mathcal{E}}_k
\end{equation}}
Now vorticity can be obtained by taking curl of the equation (\ref{mu}) and assuming that
the chemical potentials and temperature are constant in space and time:
\small{\begin{equation}
 <\boldsymbol{\omega}_k> = i\sqrt{\frac{\alpha^\prime}{\pi^3}} \frac{1}{2\nu_c}\frac{1}{\mu_R+\mu_L}(\boldsymbol{k}\times\boldsymbol{\mathcal{E}}_k)
\label{vorticity1}
\end{equation}}
One can find contribution of the vorticity to the total current from equations (\ref{sigma E}) and
$\mathcal{J}_{k\omega}^{i}= \sigma_{E}^{ij}{\mathcal{E}_k}^{j}+ \sigma_{B}^{ij}\mathcal{B}_{k}^{j}$. 
By using equation (\ref{vorticity1}) the vorticity current can be written as
\small{\begin{equation} 
\boldsymbol{\mathcal{J}_{ \omega}} \approx -\sqrt{\frac{4\pi\alpha^{\prime}}{9}}(\mu_R^2-\mu_L^2)\boldsymbol{\omega}= \xi \boldsymbol{\omega}
\label{vcurrent}
\end{equation}}
\noindent
Thus in absences of any chiral-imbalance there is no vorticity current.
Here we note that our definition agrees with the Ref.\cite{son_2009a}. In Appendix we demonstrate that 
our kinetic theory is also consistent with second law of thermodynamics.
Further using equation (\ref{vorticity1}), one can eliminating
$\left(\boldsymbol{k\times\mathcal{E}_k}\right)$ in equation (\ref{B_kw vector}) and obtain:
\small{\begin{align}
 \frac{\partial \boldsymbol{\mathcal{B}}_k}{\partial \eta}&+\frac{3\nu_c}{4\pi m_d^2}k^2\boldsymbol{\mathcal{B}}_k 
+i \frac{\sqrt{4\pi\alpha^{\prime}}\nu_c}{m_D^2} (\mu_R^2-\mu_L^2)(\boldsymbol{k}\times\boldsymbol{\omega_k}) \nonumber\\
&+i\left(\frac{4\alpha^{\prime}\nu_c }{\pi m_D^2}\right)(\mu_R-\mu_L)(\boldsymbol{k\times\mathcal{B}_k})=0.
\label{diffusivity1}
\end{align}}
In this equation, second term is usual diffusivity term, however third and fourth terms are additional term.
Which respectively represent vorticity and chiral magnetic effects on the chiral plasma. Therefore 
equation (\ref{B_kw vector}) actually contain terms due to vorticity and magnetic effect. The saturated
state of the instability can be studied by setting $\partial_{\eta} \boldsymbol{\mathcal{B}_k}=0$ in equation (\ref{diffusivity1}).
After taking a dot product of the equation (\ref{diffusivity1}) with fluid velocity $\boldsymbol{v_k}$ after
setting $\partial_{\eta} \boldsymbol{\mathcal{B}_k}=0$ one can obtain,
\small{\begin{equation}
 \left(\boldsymbol{\omega_k}-i\frac{16T\delta }{3}\boldsymbol{v}_k\right)\cdot \boldsymbol{\mathcal{B}_k}=0. \label{43}
\end{equation}}
Here we have defined $\delta= \alpha^{\prime}(\mu_R-\mu_L)/T$.
We can write expression for magnetic field, which satisfies above equation (\ref{43}) as:
\small{\begin{equation}
 \boldsymbol{\mathcal{B}_k}= g(\boldsymbol{k}) \boldsymbol{k}\times
 \left[\boldsymbol{\omega_k}-i\frac{16T \delta}{3}\boldsymbol{v}_k\right]
\end{equation}}
Where $g(k)$ is any general function, which can be determined by 
substituting the above expression for the magnetic field
 into equation (\ref{diffusivity1}) in the case of steady state. In the very large length scale i.e. $\boldsymbol{k}\rightarrow 0$:
\small{\begin{equation}
 g(\boldsymbol{k})= -\frac{3}{32}\sqrt{\frac{\pi^3}{\alpha^{\prime 3}}}\frac{\mu_R^2-\mu_L^2}{(\mu_R-\mu_L)^2}
\end{equation}}
So for a very large length scale $\boldsymbol{k}\rightarrow 0$, magnetic field in the steady state is:
\small{\begin{equation}\label{max B with g}
 \boldsymbol{\mathcal{B}_k}= -i\sqrt{\frac{\pi^3}{4\alpha^{\prime}}}\frac{\mu_R^2-\mu_L^2}{(\mu_R-\mu_L)}\boldsymbol{\omega_k}
\end{equation}}
 This equation relates the vorticity generated during the instability with the magnetic field in the steady state.

 However in the collisionless regime ($\omega\ll k$ and $\nu_c=0$), one can have an instability described by
equation (\ref{mode11}) with typical scales $k\sim \alpha^{\prime}\Delta\mu$ and $|\omega|\sim \alpha^{\prime 2} T \delta$
\cite{Akamatsu_13}. Using the expression for electric and magnetic conductivities for modes in this regime
one can write the magnetic diffusivity equation as:
\small{\begin{equation}
 \frac{\partial \boldsymbol{\mathcal{B}_k}}{\partial \eta}+ \frac{k^2}{4\pi \sigma_1}\boldsymbol{\mathcal{B}_k}- i
 \frac{T \delta}{\pi \sigma_1}\left(\boldsymbol{k}\times\boldsymbol{\mathcal{B}_k}\right)=0
\end{equation}}
where $\sigma_1 =\pi m_D^2/2k$. Here it should be noted that unlike equation (\ref{diffusivity1}), the above equation does
not have a vorticity term. 
The last term on the left-hand side arises due to the chiral-magnetic effect. 
In the steady state ($\partial_{\eta}\boldsymbol{\mathcal{B}_k}=0$), one can gets  
$\boldsymbol{\nabla\times \mathcal{B}}=(4 T\delta)\boldsymbol{\mathcal{B}}$. 
This equation resembles the case of magnetic field in a force free configuration of the conventional plasma
where the plasma pressure is assumed to be negligible in comparison with the magnetic pressure \cite{Chandrasekhar01041958}.
But for our case no such assumption about the plasma pressure is required.

\section{Results and discussion}
In the previous sections we have applied the modified kinetic theory in the presence of
chiral imbalance and obtained equations for the magnetic field generation for both the collision
dominated and the collisonless regimes. The instability can lead to generation
of the magnetic field at the cost of the chiral imbalance. This can be seen from the anomaly
equation $ n_L-n_R+2\alpha^{\prime}\mathcal{H}$=constant above $T>80$~ TeV.
Where $n_{L,R}=\frac{\mu_{L,R}T^2}{6}$ and $\mathcal{H}$ is the magnetic helicity defined as:
\begin{equation}
 \mathcal{H} =\frac{1}{V} \int d^3 x(\boldsymbol{Y}\cdot \boldsymbol{\mathcal{B}_Y})
\end{equation}
  One can estimate strength of the generated magnetic field as follows: From equations (\ref{anomaly}-\ref{CS})
one can notice that right-handed electron number density $n_R$ changes with the Chern-Simon number $n_{cs}$ of 
the hyper-charge field configuration
as $\Delta n_R =\frac{1}{2}y^2_R n_{cs}$. Here $n_{CS} \approx \frac{g^{\prime 2}}{16\pi^2} k Y^2$ and $\Delta n_R=\mu_R T^2 = 
\frac{88}{783}\delta T^3$\cite{turner}. From this, one can estimate magnitude of the generated physical magnetic field to be
\begin{equation}
        B^{phy}_Y\approx \left[\frac{\pi^2k\delta}{g^{\prime 2}\alpha^{\prime}T} \right]^{\frac{1}{2}}T^{2},
\label{Bfield}
\end{equation}
where we have used $B^{phy}_Y \sim k Y$ and $k^{-1}$ is physical length scale, which is related with the
comoving length by $k_{phy}^{-1}=(a/k_c)^{-1}$.

Now consider the regime $\omega,k\ll\nu_c$ where dynamics for the magnetic field is described by Eqs.(\ref{mode1}-\ref{mode2}).
Eq.(\ref{mode1}) clearly gives unstable modes for $\left(\frac{T \delta }{\pi m^2_D}\right)k<1$ 
is satisfied. However, Eq.(\ref{mode2}) gives a purely damping mode if the condition
$\left(\frac{T \delta }{\pi m^2_D}\right)k\ll 1$ is satisfied. One can rewrite this condition as $\left(\frac{T\delta}{3\pi\sigma\nu_c}\right)k\sim 
\left(\frac{10^{-2}}{3\pi}\frac{\delta}{\nu_c}\right)k\ll1$.Here we have used $m^2_D=3\nu_c\sigma$ with
$\nu_c\sim \alpha^{\prime 2} ln(\frac{1}{\alpha^{\prime}})T$ \cite{PhysRevD.56.5254} and $\sigma=100T$. Thus for $k\ll\nu_c$ and $ \delta \ll1$, 
equation (\ref{mode2}) can only give purely damped modes. For these values of $k$ and $\delta$,  equation (\ref{mode1}) 
assumes similar form as the equation for the magnetic field dynamics considered in Ref.\cite{Joyce_97}.
If one replaces $\frac{\partial}{\partial\eta}$ by $-i\omega$ in Eq.(\ref{mode1}), the dispersion relation
for the unstable modes can be obtained. The fastest growth of the perturbation occurs for $k_{max1}\sim \frac{8T \delta}{3}$
and the maximum growth rate can be found to be $\Gamma_1\sim \frac{16}{3\pi}\frac{T^2 \delta^2}{m^2_D}\nu_c$.
Here we note that our  $k_{max1}$  differs by a numerical factor from the value of $k$ where the peak in the magnetic 
energy calculated using chiral magnetohydrodynamics \cite{Tashiro_12}. For $\delta\sim 10^{-6}$
and $\alpha^{\prime}\sim 10^{-2}$ one can show that $\frac{k_{max1}}{\nu_c}\ll 1$ and
$\frac{\Gamma_1}{\nu_c}\ll1$ is satisfied. For these values of $k_{max1}$, $\alpha^{\prime}$ and $\delta$ one can estimate magnitude of
the generated magnetic field using equation (\ref{Bfield}). We find $B\sim 10^{26}$~Gauss for $\alpha^{\prime}\sim 10^{-2}$ and  the typical length scale
$\lambda\sim 10^5/T$. Here we would like to note that typical Hubble length scale $\sim 10^{13}/T$ which is much
larger than the typical length scale of instability. Our estimate of magnetic field strength $B$ in the collision dominated regime broadly agree with Ref.\cite{Joyce_97}.
Here we note again that equation (\ref{mode1}) includes effect of the Ohmic decay
due to presence of the collision term. Our analysis shows that Ohmic decay
is not important for the instability. Further we have shown that the chiral instability can also lead to
generation of vorticity in the collision dominated regime. Typical length scale for vorticity is similar to that of the magnetic field.
From Eq.(\ref{max B with g}) magnitude of the vorticity to be $\omega_v\sim 10^{-4}B/T$.

Next, we  analyze the chiral instability in collisionless regime $\nu_c\ll\omega\ll k$, considering
 equations (\ref{mode11}-\ref{mode22}). Here one finds the wave number $k_{max2}=\frac{8\delta T}{9}$ at which the maximum growth rate
 $\Gamma_2=\frac{1}{2\pi}\frac{T^3\delta^3}{m^2_D}$ occur. Now $\frac{k_{max2}}{\nu_c}=\frac{8}{9}\frac{\delta}{\alpha^{\prime 2}}\ll1$
 and $\frac{\Gamma_2}{\nu_c}\sim \frac{3\delta^3}{8\pi^2\alpha^3ln(1/\alpha)}\ll 1$ this puts constraints on the allowed values of
$\delta$. For $\delta\sim 10^{-1}$, $\alpha\sim 10^{-2}$ and $T\sim T_R$, one can estimate magnitude of the magnetic field to be $10^{31}$~Gauss.
Typical length scale for the magnetic field $\lambda_2\sim 10/T$ and which is much smaller
than the length scale in the collision dominated case. This is expected as the typical length scale associated with kinetic theory
are smaller than the hydrodynamical case (related with the collision dominated regime).

The upper and lower bounds on the present observed magnetic field strength from PLANCK 2015 results \cite{Ade2015}
	and blazars \cite{Neronov2010, Tavecchio01072010} are between 
	$10^{-17} G -10^{-9} G$. However, recently in Ref.\cite{PhysRevD.93.083520} it has been shown that 
	if the magnetic field is helical and created before the electroweak phase
	transition then it can produce some baryon asymmetry. This can put more stringent bounds
	on the magnetic field ($10^{-14} G -10^{-12} G$). Since the magnetic fields
	and the plasma evolutions are coupled, the produced magnetic field may not evolve
	adiabatically i.e. like $a(\eta)^{-2}$ due to the plasma processes like turbulence. Similarly the magnetic 
	correlation length $\lambda_B\propto k^{-1}_{max}$ may  not be proportional to $a(\eta)$. 
	Typical values of $\lambda_B$ for the collision dominated and and collisionless cases in our case 
	are $10^{5}/T$ and $10/T$ respectively. The length scale of turbulence can be  
	 written as $\lambda_T \approx \frac{B_p}{\sqrt{\varepsilon^{ch}+p^{ch}}}\eta
	 \sim \frac{B_p}{\sqrt{\varepsilon^{ch}+p^{ch}}}H^{-1}$, 
	 where $B_p$ is the physical value of magnetic field and $\varepsilon^{ch}$ and $p^{ch}$ are respectively
	 energy and pressure densities of the charged particles. 
	 $g_*^{ch}(T)$ and $g_*^{total}(T)$ are the number of degree of freedom of the  $U(1)$ charged particles 
	in the thermal bath. 
	For $\lambda_B\gg \lambda_T$ the effect of turbulence can be negligible. However,
	the maximum value of the magnetic field (for $\nu_c=0$) is about $10^{31}~G$ in our case and 
	this gives $\lambda_T\approx 10^{6}/T$. Thus we have  $\lambda_B\ll \lambda_T$
	and following Ref.\cite{PhysRevD.93.083520} we assume that the generated magnetic fields
	will undergo inverse cascade soon after their generation. 
	 One can relate $B_p$ and $\lambda_B$ that undergoing the process of inverse cascade with the present day
	 values of magnetic field $B_0$ and the correlation length $\lambda_0$ using the following
	 equations \cite{PhysRevD.93.083520}:
	\begin{align}
	&B_P^{IC}(T)\simeq  9.3 \times 10^9 ~G \left(\frac{T}{10^2 GeV}\right)^{7/3}\left(\frac{B_0}{10^{-14} ~G}\right)^{2/3}  \nonumber \\
	&\hspace*{4 cm} \times\left(\frac{\lambda_0}{10^2 ~pc}\right)^{1/3}\mathcal{G}_B(T)\\ 
	&\lambda_B^{IC}(T)\simeq 2.4\times 10^{-29} Mpc\left(\frac{T}{10^2 ~GeV}\right)^{-5/3}\left(\frac{B_0}{10^{-14}G}\right)^{2/3}\nonumber \\
	&\hspace*{5 cm}\times \left(\frac{\lambda_0}{1 pc}\right)\mathcal{G}_{\lambda}(T)
	\end{align}
	Where $\mathcal{G}_B(T)=(g_*^{total}(T)/106.75)^{1/6}(g_*^{ch}(T)/82.75)^{1/6}(g_{*s}\\
	(T)	/106.75)^{1/3}$ and $\mathcal{G}_{\lambda}(T)=(g_*^{total}(T)/106.75)^{-1/3} (g_*^{ch}(T)\\ 
	 /82.75)^{-1/3}(g_{*s}(T)/106.75)^{1/3}$. From these equations one can see that for collision dominated case $B_p\simeq 10^{26}~G$ 
	can be achieved when $B_0\simeq 10^{-12}~G$ and $\lambda_0\simeq 100~ Kpc$. 
	However in collisionless regime a value of  $B_0\simeq 10^{-11}~G$ and 
	$\lambda_0\simeq 1 Mpc$ at temperature $T=80 ~TeV$ gives the values that we have found in our estimates 
	for the peak value of the magnetic field. Thus the values of the magnetic field and the correlation length scale 
	estimated by us can be consistent with the current bounds obtained from CMB  observation and necessary for current observed baryon assymmetry. 
	Since the value of $B_p$ and $\lambda_B$ for the collision dominated case are simillar to that given in the 
	Ref.\cite{PhysRevD.57.2186}, so we believe that they are also consistent with BBN constraints.\\

	In conclusion we have studied generation of the magnetic field due to the anomaly in primordial plasma consisting of the standard
 model particles. We have applied the Berry curvature modified kinetic theory to study this problem.
The effect of collision in the kinetic equation was incorporated using the relaxation time approximation. We find that the chiral
instability can occur in presence of the dissipation in both collision dominated and collisionless regimes.
We find that in the collision dominated case the chiral instability can produce a magnetic field of order $10^{27}$ 
Gauss with the typical length scale $10^5/T$.  These results are in broad agreement with Ref.\cite{Joyce_97}. 
However in this work authors have used heuristic kinetic equation and the collision 
term was not explicitly written in the kinetic equation. However the expression for the total current included the Ohm's law.
We have obtained expressions for electric and magnetic conductivities using the modified kinetic theory.
We find that expression for electric conductivity in chiral plasma has a non-dissipative term in addition to the 
standard Ohmic term. It is shown that this new term is related to the vorticity current term found in the
chiral magnetohydrodynamics \cite{Giovannini:2013oga}. Further we have also studied the chiral instability 
in the collisionless regime. It is shown that in this regime magnetic field of strength $10^{31}$~Gauss can be generated at length scale
$10/T$. These length scales are much smaller than the length scale of the magnetic field in the collision dominated regime.
Further the obtained values of magnetic-field strength and the length scale are shown to be consistent
with the recent constraints from CMB data.
We have also shown that in the collision dominated regime results of kinetic theory agrees with the hydrodynamic treatment.

\section*{Appendix} 
In Ref.\cite{son_2009a} it was shown that a parity-violating hydrodynamics violate the second law
 of thermodynamics $\partial_{\alpha}s^{\alpha} \geq 0$, where $s^\alpha$is the entropy current density,
 unless certain constraints on the transport coefficients are imposed. Therefore our results in the collision
 dominated regime should be in agreement with Ref.\cite{son_2009a}.
 The most general equations for U(1) and entropy currents 
can be written as \cite{son_2009a}:
  \begin{gather}
  \nu^{\alpha}=\nu^{\prime\alpha}+\xi(\mu,T) \omega^{\alpha}
 +\xi_B(\mu,T)\mathcal{B}^{\alpha}, \label{jcovariant}\\
  s^{\alpha}=s^{\prime\alpha}+D(\mu,T)\omega^{\alpha}+D_{B}(\mu, T)\mathcal{B}^{\alpha}
  \end{gather}
  \noindent
 where, $\nu^{\prime\alpha}=\rho U^\alpha+\sigma \mathcal{E}^{\alpha}$,
 and $s^{\prime\alpha}=s U^{\alpha}-\frac{\mu}{T}\nu^{\alpha}$ with
 $\rho$ and $U^\alpha$ being the charge density and hydrodynamic four-velocity respectively. 
 Entropy density can be found using the thermodynamic relation $sT +\mu \rho = (\varepsilon+p)$ where, $\varepsilon$ 
 denotes energy density and $p$ denotes the pressure. In the collision dominated 
 (MHD) limit there is no charge separation in the plasma and one can regard total charge density as zero.
 Using the energy-momentum tensor 
 $T^{\alpha\beta}=(\varepsilon+p)u^{\alpha}u^{\beta}+pg^{\alpha\beta}$ one can write equation of motion:
 $\partial_\alpha T^{\alpha\beta}=F^{\beta\gamma}j_{\gamma}$ and divergence of the entropy current as
\small{
\begin{align} \label{entropy production}
\partial_\alpha (s^{\alpha}-D\omega^{\alpha}-&D_B B^{\alpha})=-(\nu^{\alpha}-\xi \omega^{\alpha}-\xi_B B^{\alpha})\nonumber\\
&\times \left(\partial_\alpha \frac{\mu}{T}-\frac{E_\alpha}{T}\right)-C \frac{\mu}{T}E_\alpha B^\alpha.
\end{align}}
 According to Ref.\cite{son_2009a} the second law of thermodynamic is satisfied if the following four
 equations are satisfied:
 \small{\begin{gather}
 \partial_{\alpha} D-\frac{2\partial_\alpha p}{\varepsilon +p}D-\xi\partial_\alpha\frac{\mu}{T}=0 \label{53}\\
 \partial_\alpha D_{B}-\frac{\partial_\alpha p}{\varepsilon+p}D_{B}-\xi_{B} \partial_\alpha\frac{\mu}{T}=0\label{54}\\
-2D_B+\frac{\xi}{T}=0  \label{55}\\
\frac{\xi_B}{T}-C\frac{\mu}{T}=0 \label{56}
\end{gather}}
 
 In Ref.{\cite{son_2009a}} these equations are solved and one can know dependence  of $\xi$,
 $\xi_B$, $D$ and $D_B$ on $\mu$ \& $T$ upto an arbitrary constant.
Next, we assume the perturbation scheme considered for the kinetic approach. 
 For no background field tensor, one can write $F^{\alpha\beta}=\delta F^{\alpha\beta}$.
 Also the energy density $\varepsilon$, pressure $p$ and flow velocity $U^\alpha$ can be written 
 in this scheme as: $\varepsilon=\varepsilon_0+\delta\varepsilon$, $p=p_0+\delta p$
 and $U^{\alpha}=U_0^{\alpha}+\delta u^{\alpha}$ respectively. Where all quantities with subscript $`` 0\textquotedblright$
 are background values. The background and perturbed velocities are defined as $U_0^{\alpha}=(1,0,0,0)$
 and $\delta u^{\alpha}=(0,\delta \vec{u})$ respectively. Here $\vec{u}$ is the three flow velocity.
 We assumed that background is homogeneous and isotropic.
 The equation of motion for the background gives: $\partial_0 \varepsilon_0=0$ \& $\partial_0 p_0=0$.
 Since the $\varepsilon_0$ and $p_0$ are functions of temperature and chemical-potential, we regard background
 temperature as constant.  In this scheme the zeroth and i-th components of the equation of motion can be written
 as:
\begin{eqnarray}
 \partial_0 \delta \varepsilon +(\varepsilon_0 +p_0)\partial_i \delta u^{i}=0 \label{0th}\\
 (\varepsilon_0 +p_0) \partial_0 \delta u^{i}+ g^{ij} \partial_j \delta p =0\label{ith}.
\end{eqnarray}
Ignoring the time derivative term in the Maxwell's equation one can write
$\nabla \times \delta \boldsymbol{B}=4\pi \delta \boldsymbol{j}$ and 
by using using expression for the perturbed current one can obtain the following dispersion relation 
\begin{equation}
 i\omega =\frac{k^2}{4\pi\sigma_0}\pm \frac{\xi_{0B}}{\sigma_0}k
 \label{ss09}
\end{equation}
where we have used $\vec{k}=k\hat{z}$ and $\delta\vec{B}_k= \delta\vec{B}_{kx}\hat{x}+\delta\vec{B}_{ky}\hat{y}$.
It should be noted that Eq.(\ref{ss09})
matches with the dispersion relation obtained by kinetic theory approach [equation (\ref{dispersr})].

We first emphasize that there is no current in the background and therefore the transport coefficients
 that appear in the expression for the perturbed current depends only on the background temperature
 and the chemical potentials. Now consider Eqs.(\ref{53}-\ref{56}) which for the background quantities
 can be described by the following two equations:
 \begin{gather}
 -2D_{B0}+\frac{\xi_0}{T_0}=0  \label{58}\\
\frac{\xi_{B0}}{T_0}-C\frac{\mu_0}{T_0}=0 \label{59}
\end{gather} 
These equations are satisfied by each species considered. So one can write $\xi_0=\xi_{R0}+\xi_{L0}$
and $\xi_{B0}=\xi_{BR0}+\xi_{BL0}$. Using expression for $\xi_0$ from the kinetic equation, one can calculate
$D_{B0}$ from Eq.(\ref{58}) and it agrees with the expression obtained in Ref.\cite{son_2009a} and also
the expression obtained for $\xi_B$ using kinetic theory is in agreement with it.
Thus we have shown that the modified kinetic theory respects the constraint implied by the second law of
thermodynamics.
\\ ~\\
\textbf{Acknowledgment:} 
We would like to thank our colleagues Manu George and Avadesh Kumar for their helpful comments and discussions.
We also thank anonymous referee of this work whose comments has helped us in
improving our presentation.
\bibliography{finalbib}{}
\bibliographystyle{apsrev4-1}
\end{document}